\documentclass[journal]{IEEEtran}
\usepackage{algorithm}
\usepackage{graphicx}
\usepackage{algpseudocode}
\usepackage{amsmath}
\usepackage{amssymb}
\usepackage{xcolor}
\newtheorem{thm}{Theorem}[section]

\newtheorem{lem}{Lemma}[section]
\usepackage{multirow}
\newtheorem{defn}{Definition}[section]
\usepackage{lineno,hyperref}
\usepackage{array}
\usepackage[tight,normalsize,sf,SF]{subfigure}
\newcommand*{\QEDB}{\hfill\ensuremath{\square}}

\ifCLASSINFOpdf
\else
\fi
\begin{document}
%
\title{A generalized power iteration method for solving quadratic problem on the Stiefel manifold}
%
%
%

\author{Feiping Nie, Rui Zhang, and Xuelong Li,\IEEEmembership{~Fellow~IEEE}
\thanks{Feiping Nie and Rui Zhang are with School of Computer Science and Center
for OPTical IMagery Analysis and Learning (OPTIMAL), Northwestern Polytechnical University, Xi'an
710072, Shaanxi, P. R. China.}
\thanks{E-mail: \{ruizhang8633, feipingnie\}@gmail.com.}
\thanks{Xuelong Li is with Center for OPTical IMagery Analysis and Learning (OPTIMAL), State Key Laboratory of Transient Optics and Photonics, Xi'an Institute of Optics and Precision Mechanics, Chinese Academy of Sciences, Xi'an 710119, Shaanxi, P. R. China. }
\thanks{E-mail: xuelong$\_$li@ieee.org.}}

\maketitle

\begin{abstract}
In this paper, we first propose a novel generalized power iteration method (GPI) to solve the quadratic problem on the Stiefel manifold (QPSM) as $\min_{W^TW=I}$ $Tr(W^TAW-2W^TB)$ along with the theoretical analysis. Accordingly, its special case known as the orthogonal least square regression (OLSR) is under further investigation.  Based on the aforementioned studies,  we then cast major focus on solving the unbalanced orthogonal procrustes problem (UOPP).  As a result, not only a general convergent algorithm is derived theoretically but the efficiency of the proposed approach is verified empirically as well.
\end{abstract}

\begin{IEEEkeywords}
quadratic problem, Stiefel manifold, power iteration, procrustes problem, orthogonal least square regression.
\end{IEEEkeywords}

%
\IEEEpeerreviewmaketitle

\section{Introduction}
The orthogonal procrustes problem (OPP) is the least square problem on the Stiefel manifold. The OPP originates from the factor analysis in psychometrics during 1950s and 1960s \cite{G,H.C}. 	The major purpose is to determine an orthogonal matrix that rotates the factor matrix to best fit some hypothesis matrix. The balanced case of  the OPP was surveyed in multiple introductory textbooks such as \cite{G.L,T}.

\begin{table*}[htbp]
\centering
\caption{Orders of complexity for 6 algorithms.}\label{tab2.5}
\begin{tabular}{cccc}
\multicolumn{4}{c}{($t$ stands for the iteration number and $(n,m,k)$  stands for the dimension.)  }\\
\hline
&RSR \cite{P}&LSR \cite{B.L}&SP \cite{Z.D}  \\ \cline{1-4}
Order of the&\multirow{2}{*}{$O(mnk+m^3kt)$}&\multirow{2}{*}{$O(mnk+m^3kt)$}&\multirow{2}{*}{$O(mnk+(m^2n+m^3)t)$}\\ 
complexity&&&\\
\hline
&LR \cite{X.H}&EB \cite{G.G}&GPI (our)\\ \cline{1-4}
Order of the&\multirow{2}{*}{$O(m^2n+nk^2+m^2kt)$}&\multirow{2}{*}{$O(m^3+(m^2n+m^3)t)$}& \multirow{2}{*}{$O(m^2n+m^2kt)$} \\
complexity&&&\\
\hline
\end{tabular}
\end{table*}

Recently,  due to the wide applications of the orthogonal regression in computer science, see \cite{R.S.C.R,C.T}, solving the unbalanced OPP  (UOPP) is under increasing concern. Multiple approaches are proposed to solve UOPP such as the expansion balanced algorithm (EB), the right hand side and the left hand side relaxation (RSR), (LSR),  the successive projection (SP)  and the Lagrangian relaxation (LR).  In \cite{G.G}, the EB method employs the expanded balanced OPP as its objective function.  In \cite{P} and \cite{B.L} respectively, the RSR and the LSR approaches update the solution row by row or column by column iteratively based on solving the least square regression with a quadratic equality 
constraint (LSQE). In \cite{Z.D}, the SP method updates the solution column by column by virtue of the projection method combined with correction techniques (PMCT) discussed by \cite{Z.H}, which is efficient to solve LSQE. In \cite{X.H},  the LR method solves UOPP by selecting different  Lagrangian multipliers. 

All the approaches mentioned above could  converge to the solution of UOPP successfully, whereas they deal with more complex procedures, which represent high orders of complexity.  Furthermore, all these methods initialize the parameters deliberately to optimize their proposed algorithms. Last but not least, all these approaches are unable to deal with a more general problem known as the quadratic problem on the Stiefel manifold (QPSM).

To address the referred deficiencies, we derive a novel generalized power iteration method (GPI) for QPSM in order to efficiently solve the orthogonal least square regression (OLSR) and UOPP with a random initial guess and concise computational steps. In sum, the proposed GPI method can deal with a more general problem known as QPSM than other approaches. Furthermore, the experimental results show that the proposed GPI method not only takes much less CPU time for the convergence but becomes more efficient dealing with the data matrix of large dimension as well.

\textbf{Notations:} For any matrix $M$, Frobenius norm is defined as $\Vert M\Vert_F^2=Tr(M^TM)$, where $Tr(\cdot)$ is the trace operator.  For any positive integer $n$, $I_n$ denotes a $n\times n$ identity matrix.

\section{Power Iteration Method Revisited}
The power iteration method is an iterative algorithm to seek the dominant eigenvalue and the related eigenvector of any given symmetric matrix $A\in\mathbb{R}^{m\times m}$, where the dominant eigenvalue is defined as the greatest eigenvalue in magnitude. The power iteration can be performed as the following steps:

1. Initialization. Random initialize a vector $w\in\mathbb{R}^{m\times 1}$, which has a nonzero 

component in the direction of the dominant eigenvector.

2. Update $m\leftarrow Aw$.

3. Calculate $q=\frac{m}{\Vert m\Vert_2}$.

4. Update  $w\leftarrow q$.

5. Iteratively perform the step 2-4 until convergence.
~\\
The power iteration could be further extended to the orthogonal iteration (also called subspace iteration or simultaneous iteration) method to find the first $k\ (k\leq m)$ dominant eigenvalues and their associated eigenvectors for the given matrix $A$. The orthogonal iteration method could be described as the following iterative algorithm:

1. Initialization.  Random initialize $W\in\mathbb{R}^{m\times k}$.

2.  Update $M \leftarrow AW$.

3. Calculate $QR=M$ via the compact QR factorization of 

$M$, where $Q\in\mathbb{R}^{m\times k}$ and $R\in\mathbb{R}^{k\times k}$.

4. Update $W\leftarrow Q$.
 
5. Iteratively perform the step 2-4 until convergence.
~\\
Apparently, the orthogonal iteration method above indicates a normalization process, which is similar as the normalization in the  power iteration method. When the matrix $A$ is positive semi-definite (psd), the orthogonal iteration method is equivalent to solving the following optimization problem 
\begin{equation}
\max_{W^TW=I_k}Tr(W^TAW). \label{e000}
\end{equation}
Therefore, the orthogonal iteration method is equivalent to the following steps under the psd matrix $A$:

1. Initialization.  Random initialize $W\in\mathbb{R}^{m\times k}$.

2. Update $M\leftarrow AW$.

3. Calculate $USV^T=M$ via the compact SVD method of $M$, 

where $U\in\mathbb{R}^{m\times k}$, $S\in\mathbb{R}^{k\times k}$ and $V\in\mathbb{R}^{k\times k}$.

4. Update $W\leftarrow UV^T$.

5. Iteratively perform the step 2-4 until convergence.
~\\
From the observation, the solution of the above algorithm as $WK$ differs from the solution of the orthogonal iteration method as $W$ by the form, where $KK^T=I_k$. However, the difference between the solutions of these two algorithms doesn't affect the objective value of the problem (\ref{e000}) due to the following derivation
$$Tr((WK)^TAWK)=Tr(W^TAWKK^T)=Tr(W^TAW).$$

\section {Quadratic Problem on the Stiefel Manifold}
The Stiefel manifold $\nu_{m,k}$ is a set of the matrices $W\in\mathbb{R}^{m\times k}$, which have orthonormal columns as
$\nu_{m,k}=\{W\in\mathbb{R}^{m\times k}: W^TW=I_k\}$.

In this section,  a novel approach  is derived  to unravel the following quadratic problem on the Stiefel manifold (QPSM) \cite{MYPR} as
\begin{equation}
\min_{W^TW=I_k}Tr(W^TAW-2W^TB)\label{e01}
\end{equation}
where $W\in\mathbb{R}^{m\times k}$, $ B\in\mathbb{R}^{m\times k}$ and the symmetric matrix $A\in\mathbb{R}^{m\times m}$.
In order to solve the problem (\ref{e01}),  
QPSM in  (\ref{e01}) can be further relaxed into 
\begin{equation}
\max_{W^TW=I_k}Tr(W^T\Tilde{A} W)+2Tr(W^TB)\label{e02}
\end{equation}
where $\Tilde{A}=\alpha I_m-A\in\mathbb{R}^{m\times m}$.  The relaxation parameter $\alpha$ is an arbitrary constant such that $\Tilde{A}$  is a positive definite (pd) matrix. 
Instead of the method of the Lagrangian multipliers to deal with an optimization problem with orthogonal constraints, one may use
a geometric optimization algorithm tailored to the Stiefel manifold, such as, for example, the one surveyed in \cite{fiori1}.

Accordingly, the Lagrangian function for the problem (\ref{e02}) can be written as
\begin{equation}
L_1(W,\Lambda)=Tr(W^T\Tilde{A} W)+2Tr(W^TB)-Tr(\Lambda(W^TW-I_k)).\label{e09}
 \end{equation}

 From Eq. (\ref{e09}), we could obtain the KKT condition for the problem (\ref{e02}) as
  \begin{equation}
\frac{\partial L_1}{\partial W}=2\Tilde{A}W+2B-2W\Lambda=0\label{e010}
 \end{equation}
~\\
which is difficult to solve directly. Thus, motivated by \cite{N.Y.H} and the power iteration method mentioned in Section 2,  we could propose the following iterative algorithm: 

1. Initialization. Random initialize $W\in\mathbb{R}^{m\times k}$ such that $W^TW=I_k$.

2. Update $M\in\mathbb{R}^{m\times k}\leftarrow 2\Tilde{A}W+2B$.

3. Calculate $W^*$ by solving the following problem
\begin{equation}
\max_{W^TW=I_k}Tr(W^TM). \label{e05}
\end{equation}

4. Update $W\leftarrow W^*$.

5. Iteratively perform the step 2-4 until convergence.
~\\
Besides, a closed form solution of the problem (\ref{e05}) can be achieved by the following derivation. 

Suppose the full SVD of $M$ is $M=\mathbb{U}\Sigma \mathbb{V}^T$ with $\mathbb{U}\in\mathbb{R}^{m\times m}$, $\Sigma\in\mathbb{R}^{m\times k}$ and $\mathbb{V}\in\mathbb{R}^{k\times k}$, then we have 
\begin{equation*}
\begin{aligned}
Tr(W^TM)&=Tr(W^T\mathbb{U}\Sigma \mathbb{V}^T)\\
&=Tr(\Sigma \mathbb{V}^TW^T\mathbb{U})\\
&=Tr(\Sigma Z)=\sum_{i=1}^k\sigma_{ii}z_{ii}
\end{aligned}
\end{equation*}
where $Z=\mathbb{V}^TW^T\mathbb{U}\in\mathbb{R}^{ k\times m}$ with $z_{ii}$ and $\sigma_{ii}$ being the $(i,i)$-th elements of the  matrix $Z$ and $\Sigma$, respectively. 

 Note that $ZZ^T=I_k$,  thus $\vert z_{ii}\vert\leq 1$.  On the other hand, $\sigma_{ii}\geq 0$ since $\sigma_{ii}$ is a singular value of the matrix $M$. Therefore, we have 
\begin{equation*}
Tr(W^TM)=\sum_{i=1}^kz_{ii}\sigma_{ii}\leq\sum_{i=1}^k\sigma_{ii}.
\end{equation*}

Apparently, the equality holds when $z_{ii}=1, (1\leq i \leq k)$. That is to say, $Tr(W^TM)$ reaches the maximum when the matrix $Z=[I_k,0]\in\mathbb{R}^{ k\times m}$.  Recall that $Z=\mathbb{V}^TW^T\mathbb{U}$, thus the optimal solution to the problem (\ref{e05}) can be represented as
\begin{equation}
W=\mathbb{U}Z^T\mathbb{V}^T=\mathbb{U}[I_k;0]\mathbb{V}^T.\label{eq1}
\end{equation}
Since Eq. (\ref{eq1}) is based upon the full SVD of the matrix $M$,  Eq. (\ref{eq1})  can be rewritten as  $W=UV^T$ via the compact SVD of the matrix $M$, where $M=USV^T$ with $U\in\mathbb{R}^{m\times k}$, $S\in\mathbb{R}^{k\times k}$ and $V\in\mathbb{R}^{k\times k}$.
\begin{figure*}[htb]
    \centering
  
    \begin{tabular}{m{0.32\textwidth}m{0.32\textwidth}m{0.4\textwidth}}
    \subfigure[$(50,100,30)$]{\includegraphics[width=0.32\textwidth,height=0.28\textwidth]{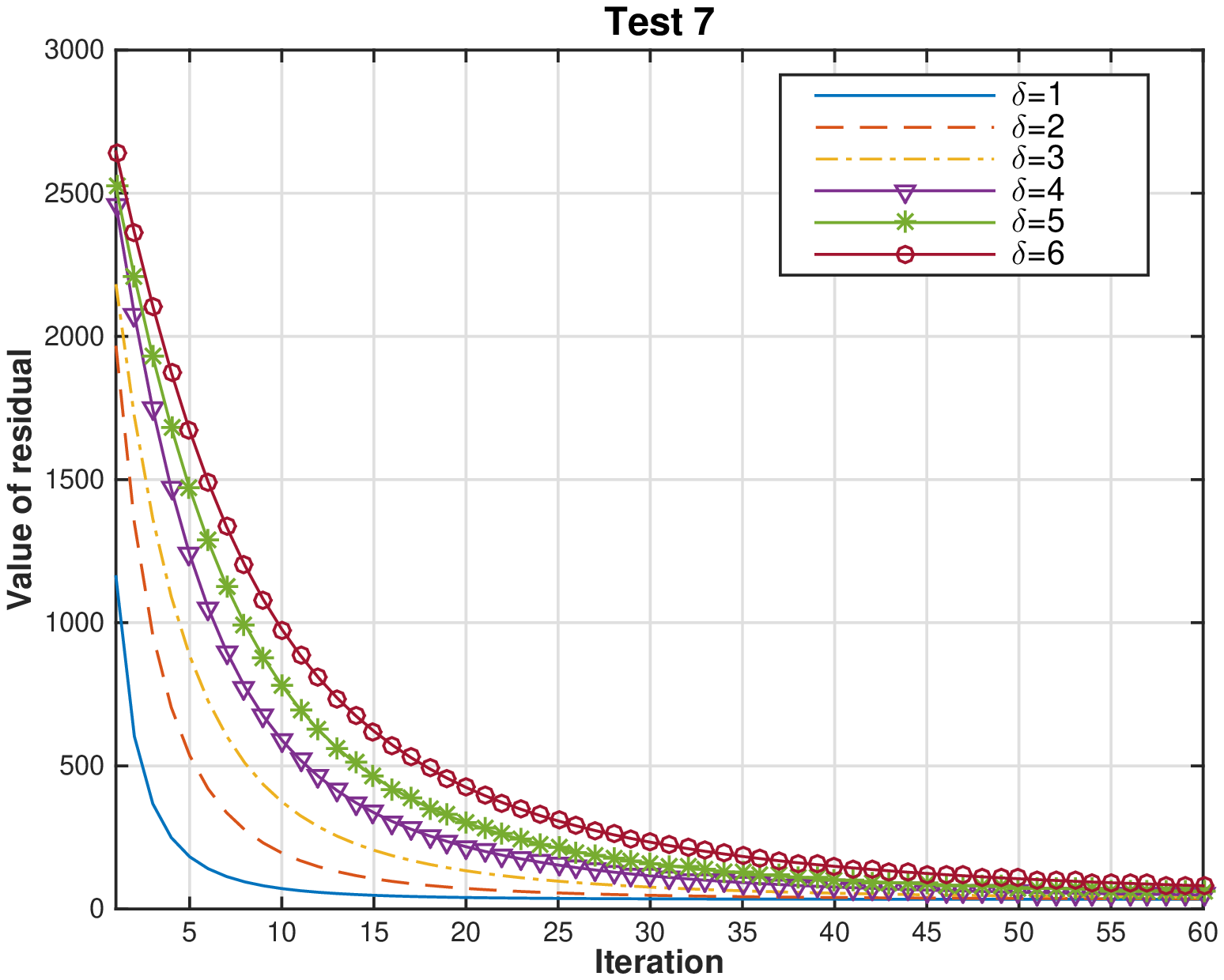}}
    &\subfigure[$(80,170,80)$]{\includegraphics[width=0.32\textwidth,height=0.28\textwidth]{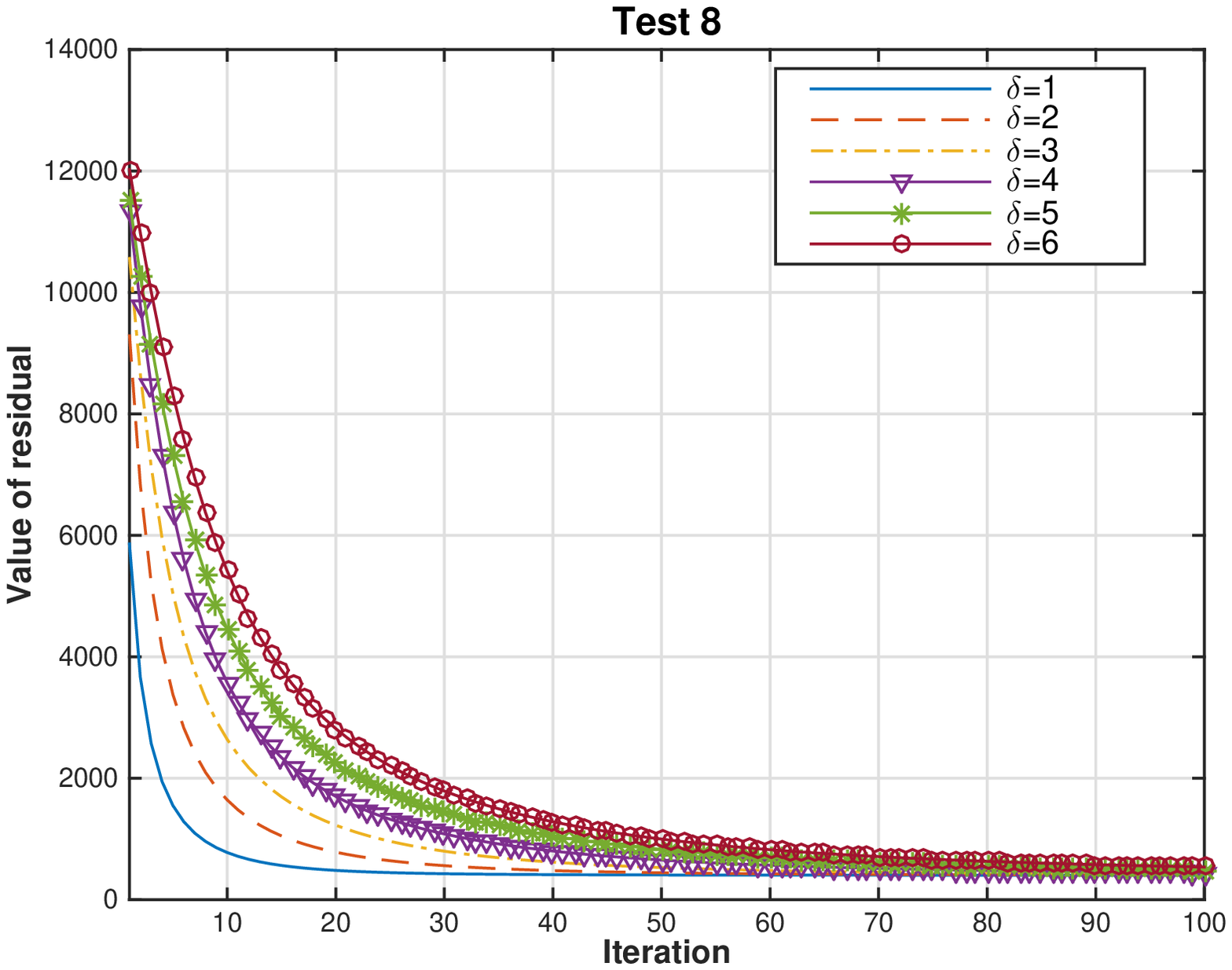}}&\subfigure[$(40,120,60)$]{\includegraphics[width=0.32\textwidth,height=0.28\textwidth]{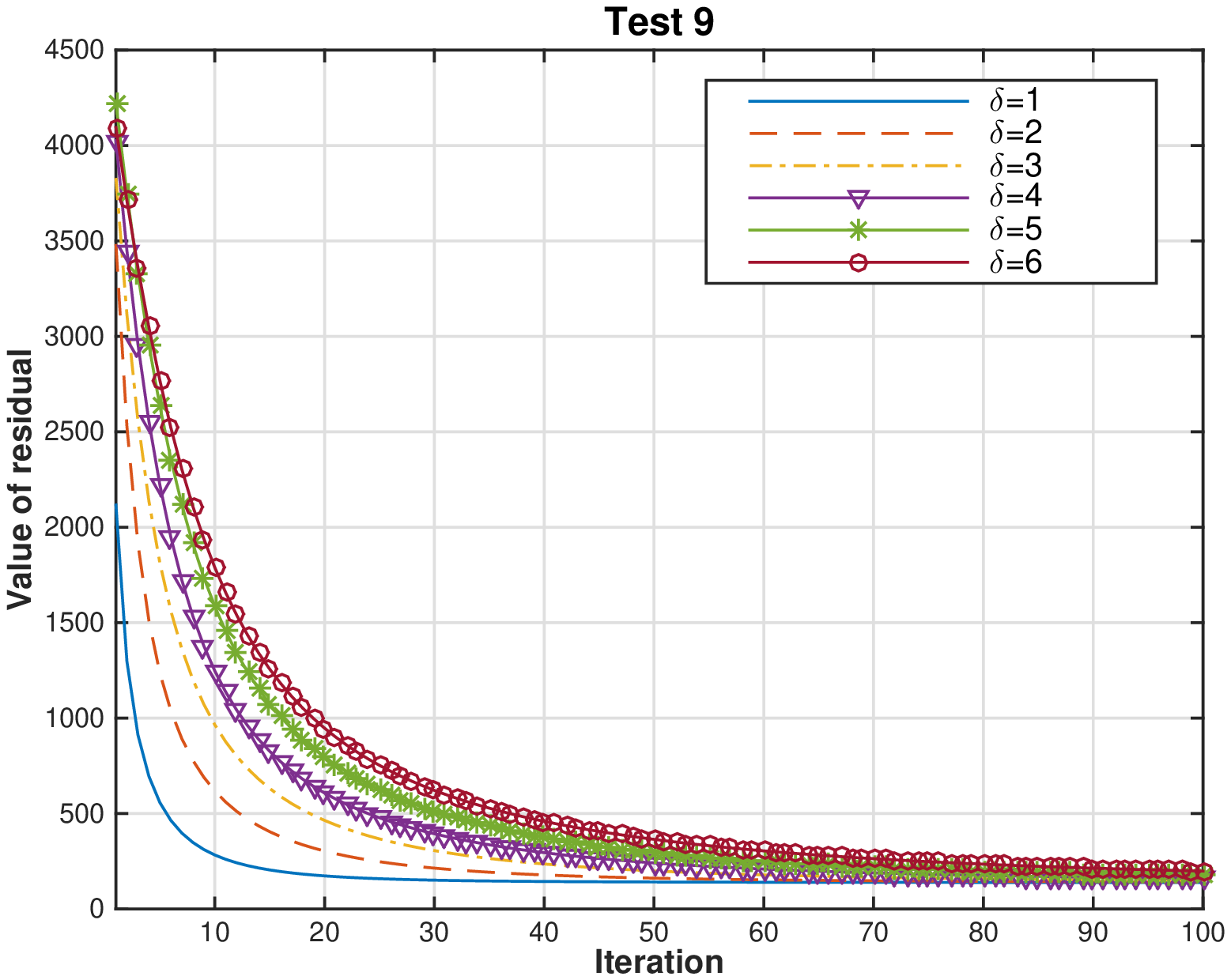}}\\
   \end{tabular}
  \caption{Comparisons of 6 different values of $\delta$ are performed under the GPI method with 3 different data matrices.}\label{fig3}
\end{figure*}

Based on the above analysis, the generalized power iteration method (GPI) can be summarized in the algorithm \ref{alg1}. 

We will prove that the proposed algorithm \ref{alg1} converges monotonically to the local minimum of QPSM (\ref{e01}).
\begin{algorithm}[htb]
\caption{Generalized power iteration method (GPI)}
\label{alg1}
\begin{algorithmic}[1]
\State \textbf{Input:}  The symmetric matrix $A\in\mathbb{R}^{m\times m}$ and the matrix $B\in\mathbb{R}^{m\times k}$.
\State  \textbf{Initialize} a random $W\in\mathbb{R}^{m\times k}$ satisfying $W^TW=I_k$ and $\alpha$ such that $\Tilde{A}=\alpha I_m-A\in\mathbb{R}^{m\times m}$ is a positive definite matrix.
\State Update $M\leftarrow 2\Tilde{A}W+2B$.
\State Calculate $USV^T=M$ via the compact SVD method of $M$ where $U\in\mathbb{R}^{m\times k}$, $S\in\mathbb{R}^{k\times k}$ and $V\in\mathbb{R}^{k\times k}$. 
\State Update $W\leftarrow UV^T$.
\State Iteratively perform the step 3-5 until the algorithm converges. 
\end{algorithmic}
\end{algorithm}

Step 5 of algorithm \ref{alg1} is an instance of a class of methods, called manifold retractions,
to update a matrix on the Stiefel manifold, that were discussed in details in
 \cite{fiori2}.

\subsection{Theoretical analysis of GPI}
\begin{lem}
If the symmetric matrix $\Tilde{A}\in\mathbb{R}^{m\times m}$ is positive definite (pd), then 
\begin{equation*}
Tr(\tilde{W}^T\Tilde{A}\tilde{W})-2Tr(\tilde{W}^T\Tilde{A}W)+Tr(W^T\Tilde{A}W)\geq 0
\end{equation*} 
where $\tilde{W}\in\mathbb{R}^{m\times k}$ and $W\in\mathbb{R}^{m\times k}$ are arbitrary matrices.\label{lem1}
\end{lem}

Proof: Since the matrix $\Tilde{A}$ is positive definite (pd), we could rewrite $\Tilde{A}=L^TL$ via Cholesky factorization.  Therefore, we have the following proof for Lemma \ref{lem1} as
\begin{equation*}
\begin{aligned}
&\Vert L\tilde{W}-LW \Vert_F^2\geq 0 \\
&\Rightarrow Tr(\tilde{W}^T\Tilde{A}\tilde{W})-2Tr(\tilde{W}^T\Tilde{A}W)+Tr(W^T\Tilde{A}W)\geq 0
\end{aligned}
\end{equation*}
\QEDB

\begin{thm}
 The algorithm \ref{alg1} decreases the value of the objective function in (\ref{e01}) monotonically in each iteration until it converges.   \label{thm2}
\end{thm}

Proof: Suppose the updated $W$ is $\tilde{W}$  in the algorithm \ref{alg1}, then we have   
\begin{equation}
Tr(\tilde{W}^TM)\geq Tr(W^TM)\label{e06}
 \end{equation}
since $\tilde{W}$ is the optimal solution of the problem (\ref{e05}). Based on the fact that $M=2\Tilde{A}W+2B$, Eq. (\ref{e06}) can be further illustrated as   
\begin{equation}
2Tr(\tilde{W}^T\Tilde{A}W)+2Tr(\tilde{W}^TB)\geq 2Tr(W^T\Tilde{A}W)+2Tr(W^TB).\label{e07}
 \end{equation}

Based on Lemma \ref{lem1} and Eq. (\ref{e07}),  we could infer that 
 \begin{equation*}
\begin{aligned}
&Tr(\tilde{W}^T\Tilde{A}\tilde{W})+2Tr(\tilde{W}^TB)\geq Tr(W^T\Tilde{A}W)+2Tr(W^TB)\\
&\Rightarrow Tr(\tilde{W}^TA\tilde{W})-2Tr(\tilde{W}^TB)\leq Tr(W^TAW)-2Tr(W^TB) 
\end{aligned}
\end{equation*}
which indicates that the algorithm \ref{alg1}  decreases the objective value of QPSM in (\ref{e01}) in each iteration  until the algorithm converges. \QEDB

\begin{thm}
 The algorithm \ref{alg1} converges to a local minimum of the QPSM problem (\ref{e01}). 
\end{thm}
Proof:  Since the algorithm \ref{alg1} performs based on solving the problem (\ref{e05}) in each iteration, the Lagrangian function for the solution of the algorithm \ref{alg1} can be represented as   
 \begin{equation}
L_2(W,\Lambda)=Tr(W^TM)-Tr(\Lambda(W^TW-I_k)).\label{e011}
 \end{equation}
 Therefore, the solution of the algorithm \ref{alg1} satisfies the following KKT condition
\begin{equation}
\frac{\partial L_2}{\partial W}=M-2W\Lambda=0\label{e012}
 \end{equation}

Generally speaking, the matrix $M$ will be updated by $\tilde{W}$ in each iteration under the algorithm \ref{alg1}. Since the algorithm \ref{alg1} converges to the optimal solution $W$ i.e. $\tilde{W}=W$ due to Theorem \ref{thm2}, Eq. (\ref{e012}) can be further formulated by substituting $M=2\Tilde{A}W+2B$ as 
\begin{equation}
\frac{\partial L_2}{\partial W}=2\Tilde{A}W+2B-2W\Lambda=0.\label{e013}
 \end{equation}
By comparing Eq. (\ref{e010}) and (\ref{e013}), we could draw the conclusion that the solution of the algorithm \ref{alg1} and the problem (\ref{e02})  satisfy the same KKT condition. 

Therefore, the algorithm \ref{alg1} converges to a local minimum of QPSM (\ref{e01}) since the problems (\ref{e01}) and  (\ref{e02}) are equivalent. \QEDB
\begin{figure*}[htb]
    \centering
  
    \begin{tabular}{m{0.32\textwidth}m{0.32\textwidth}m{0.4\textwidth}}
    \subfigure[$(100,10,100)$]{\includegraphics[width=0.32\textwidth,height=0.27\textwidth]{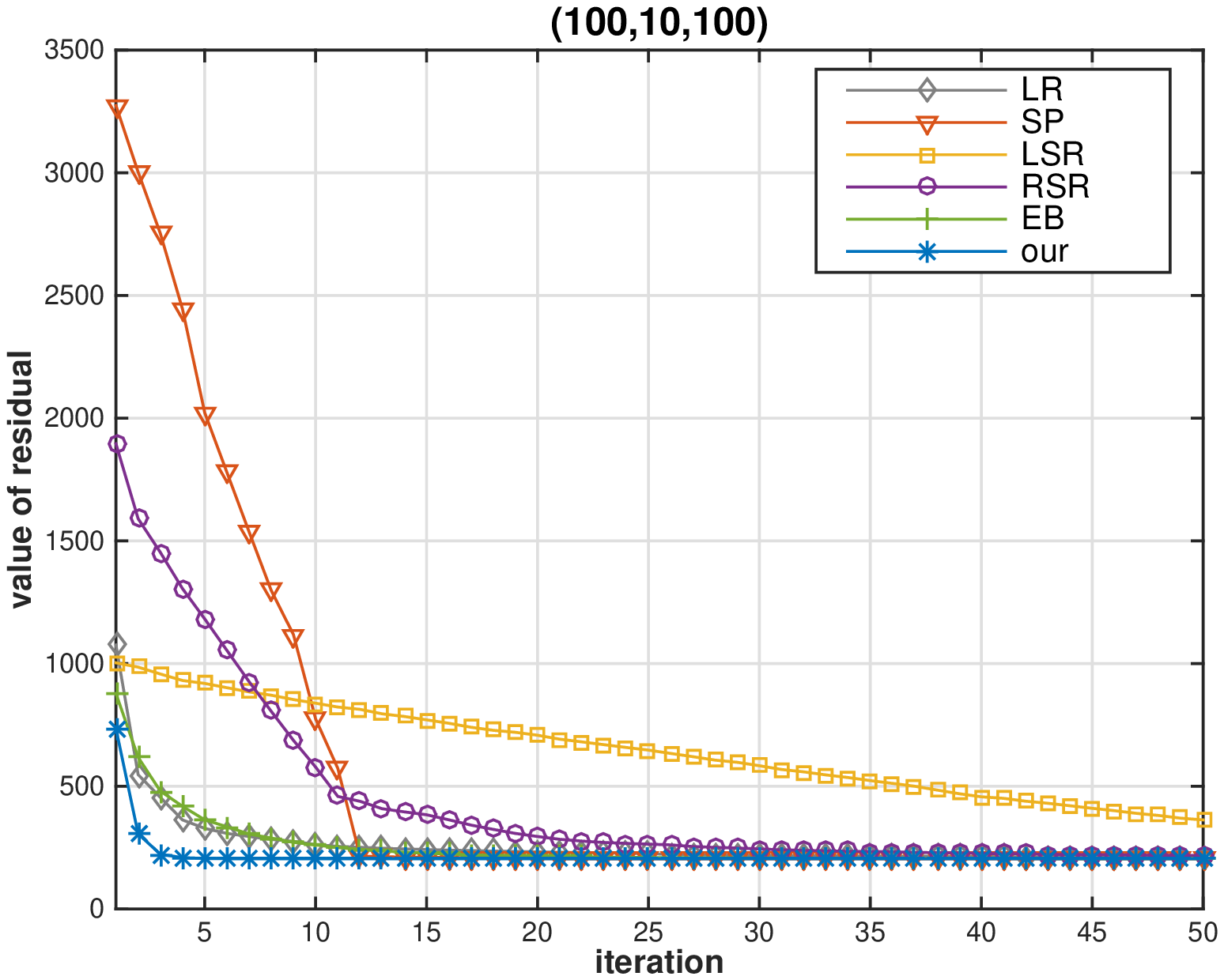}}
    &\subfigure[$(100,15,100)$]{\includegraphics[width=0.32\textwidth,height=0.27\textwidth]{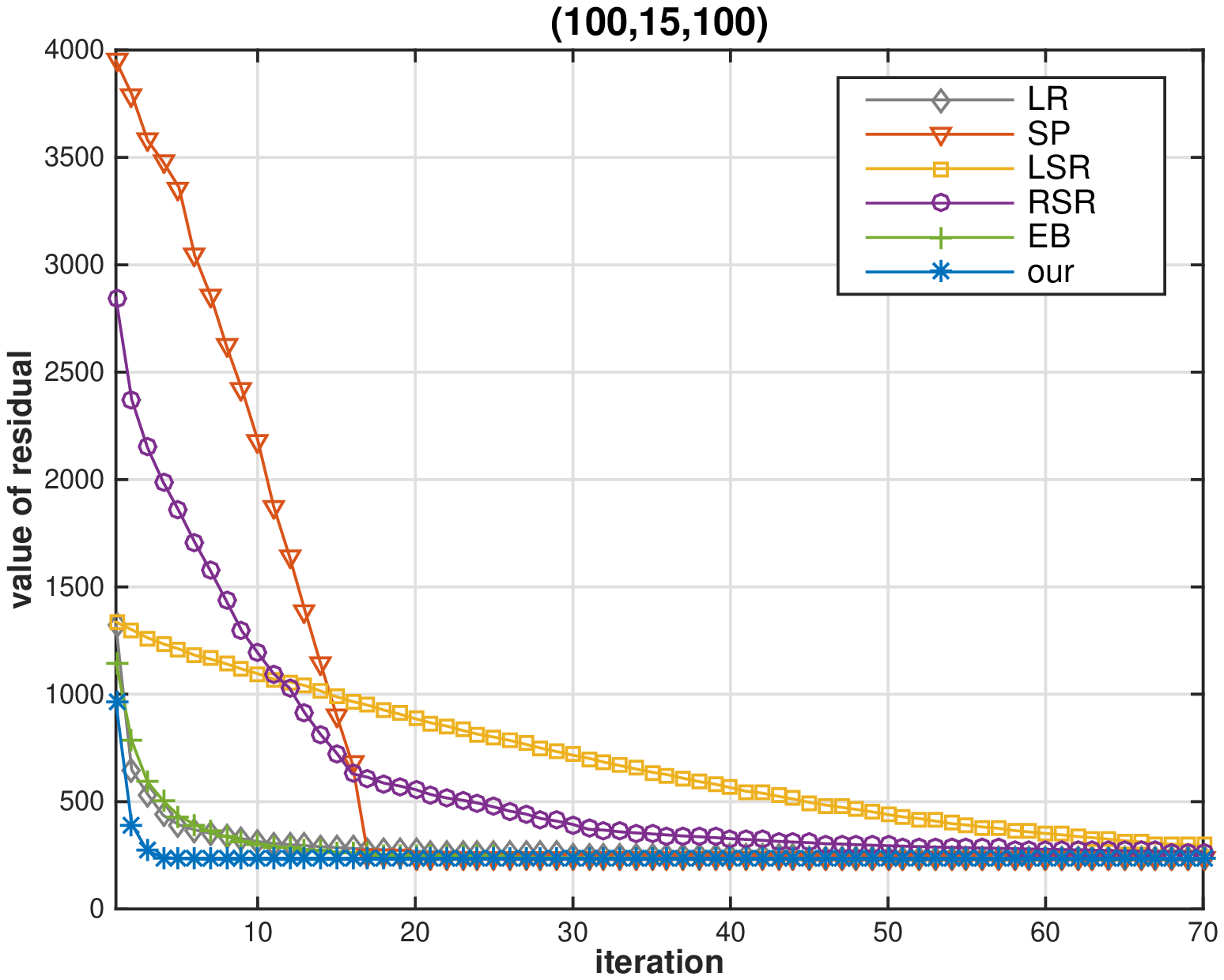}}&\subfigure[$(200,15,200)$]{\includegraphics[width=0.32\textwidth,height=0.27\textwidth]{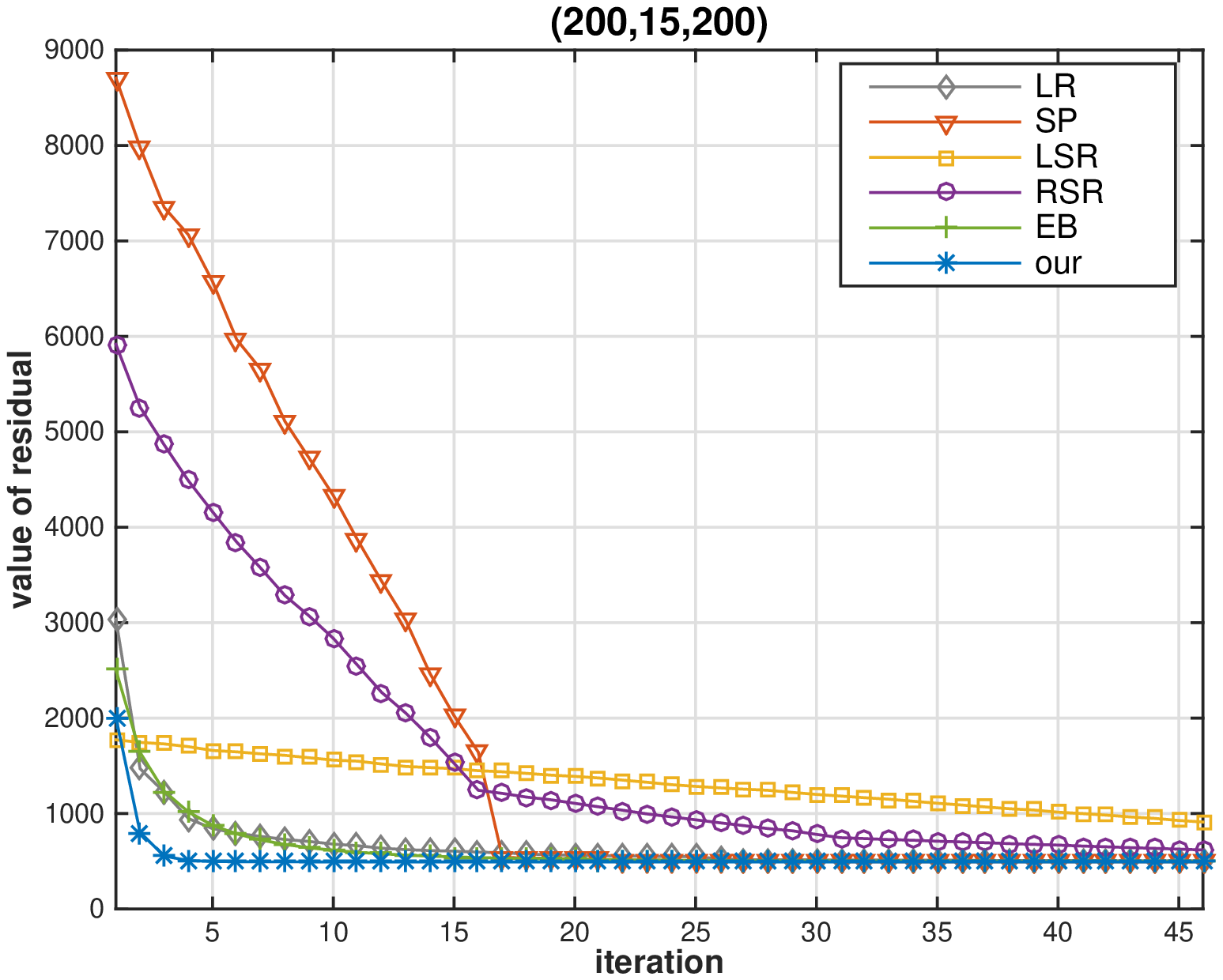}}\\
   \end{tabular}
  \caption{Comparisons of the convergence rate are performed for 6 approaches including EB \cite{G.G}, RSR \cite{P}, LSR \cite{B.L}, SP \cite{Z.D} LR \cite{X.H} and our proposed GPI method under 3 different data matrices.}\label{fig4}
\end{figure*}

Besides, the problem (\ref{e05})  has an unique solution under full column-rank matrix $M$ due to the uniqueness of the SVD method.  On the other hand, the experimental results in Section 5 represent that the proposed GPI method uniformly converges to the same objective value with a large amount of random initial guesses.  Based on the unique solution of the problem  (\ref{e05}) and the associated experimental results, it is rational to conjecture that the proposed GPI method converges to the global minimum of QPSM.

\section{Two Special Cases of Quadratic Problem on the Stiefel Manifold}  
\subsection{Orthogonal Least Square Regression}
The orthogonal least square regression (OLSR) can be written as 
\begin{equation}
\min_{W^TW=I_k, b}\Vert X^TW+\textbf{1}b^T-Y\Vert_F^2 \label{e00}
\end{equation}
where the data matrix $X\in\mathbb{R}^{m\times n}$ and the hypothesis matrix $Y\in\mathbb{R}^{n\times k}$ with $\textbf{1}=(1,1,\cdots,1)^T\in\mathbb{R}^{n\times 1}$. Moreover,  $W\in\mathbb{R}^{m\times k}$ is the regression matrix and  $b\in\mathbb{R}^{k\times 1}$ is the bias vector.  Obviously, $b$ is free from any constraint. By virtue of the extreme value condition w.r.t. $b$,  we can derive as
\begin{equation*}
\begin{aligned}
&\frac{\partial\Vert X^TW+\textbf{1}b^T-Y\Vert_F^2}{\partial b}=0\\
&\Rightarrow W^TX\textbf{1}+b\textbf{1}^T\textbf{1}-Y^T\textbf{1}=0\\
&\Rightarrow b=\frac{1}{n}(Y^T\textbf{1}-W^TX\textbf{1}).
\end{aligned}
\end{equation*}

By substituting the above result as $b=\frac{1}{n}(Y^T\textbf{1}-W^TX\textbf{1})$, Eq. (\ref{e00}) can be simplified to the following form as
\begin{equation}
\min_{W^TW=I_k}\Vert H(X^TW-Y)\Vert_F^2\label{eee}
\end{equation}
where $H=I_n-\frac{1}{n}\textbf{1}\textbf{1}^T$.

Accordingly, the problem (\ref{eee}) can be further reformulated into 
\begin{equation}
\min_{W^TW=I_k}Tr(W^TAW-2W^TB) \label{eeee}
\end{equation}
in which
\begin{equation*}
\left\{
\begin{aligned}
A&=XHX^T \\
B&=XHY\\
\end{aligned}
\right .   .
\end{equation*}
Apparently,  Eq. (\ref{eeee}) is in the exact same form as QPSM in (\ref{e01}).
Therefore,  OLSR in (\ref{e00}) can be solved via the algorithm \ref{alg1}.
\subsection {Unbalanced Orthogonal Procrustes Problem}
\begin{defn}
    With $	Q\in\mathbb{R}^{m\times k}$, $E\in\mathbb{R}^{n\times m}$ and $G\in\mathbb{R}^{n\times k}$, we
 name the optimization problem  
 


\begin{equation}
\min_{Q^TQ=I_k}\Vert EQ-G\Vert_F^2  \label{e1}
\end{equation}

1. balanced orthogonal procrustes problem (OPP) if and only if $m=k$.  
  
2. unbalanced orthogonal procrustes problem (UOPP) if and only if $m>k$. \label{1}
Especially when $Q$ serves as a column vector $(k=1)$, the problem (\ref{e1}) degenerates to  
\begin{equation}
\min_{q^Tq=1}\Vert Eq-g\Vert_2^2\label{ls}
\end{equation}
which is known as the least square problem with a quadratic equality constraint (LSQE).
 \end{defn}
 
\subsubsection{Balanced orthogonal procrustes problem revisited}

To solve the balanced OPP ($m=k$), we could expand Eq. (\ref{e1}) into
\begin{equation*}
\begin{aligned}
&\min_{Q^TQ=I_k}\Vert EQ-G\Vert_F^2\\
&\Rightarrow \min_{Q^TQ=I_k}\Vert E\Vert_F^2+\Vert G\Vert_F^2-2Tr(Q^TE^TG)\\
&\Rightarrow\max_{Q^TQ=I_k}Tr(Q^TE^TG)
\end{aligned}
\end{equation*}
~\\
which is same as the problem (\ref{e05}) with treating $E^TG=M$. 

Thus, the balanced OPP has the analytical solution of the closed form (\ref{eq1}).
\subsubsection{Unbalanced orthogonal procrustes problem}

When $m>k$, UOPP (\ref{e1}) can be expanded into
\begin{equation}
\begin{aligned}
&\min_{Q^TQ=I_k}\Vert EQ-G\Vert_F^2\\
&\Rightarrow\min_{Q^TQ=I_k}Tr(Q^TE^TEQ-2Q^TE^TG).
\end{aligned}\label{e3}
\end{equation}

Denote $E^TE=A$ and $E^TG=B$, then Eq. (\ref{e3}) is in the exact same form as QPSM (\ref{e01}). Based on the algorithm \ref{alg1}, the algorithm \ref{alg2} can be proposed to converge to a local minimum of UOPP monotonically due to the theoretical supports proved in Section 3. 

 \begin{algorithm}[htb]
\caption{GPI for solving UOPP in (\ref{e1})}
\begin{algorithmic}[1]
\State \textbf{Input:}  The matrix $E\in\mathbb{R}^{n\times m}$ and the matrix $G\in\mathbb{R}^{n\times k}$ where $m>k$.
\State \textbf{Initialize} $Q\in\mathbb{R}^{m\times k}$ and $\gamma$ such that $Q^TQ=I_k$ and the matrix $\gamma I_m-E^TE$ is positive definite, respectively. 

\State \textbf{While} not converge \textbf{do}
\State Update matrix $M\leftarrow 2(\gamma I_m-E^TE)Q+2E^TG$.
\State Calculate $U\in\mathbb{R}^{m\times k}$ and $V\in\mathbb{R}^{ k\times k}$ via the compact SVD of $M$ as $M=USV^T$.

\State Update $Q\leftarrow UV^T$.
 
\State \textbf{End while}
\State \textbf{Return} $Q$.
\end{algorithmic}
\label{alg2}
\end{algorithm}

\begin{figure*}[htb]
    \centering
  
    \begin{tabular}{m{0.45\textwidth}m{0.5\textwidth}}
        \subfigure[$(900,1000)$]{\includegraphics[width=0.5\textwidth,height=0.33\textwidth]{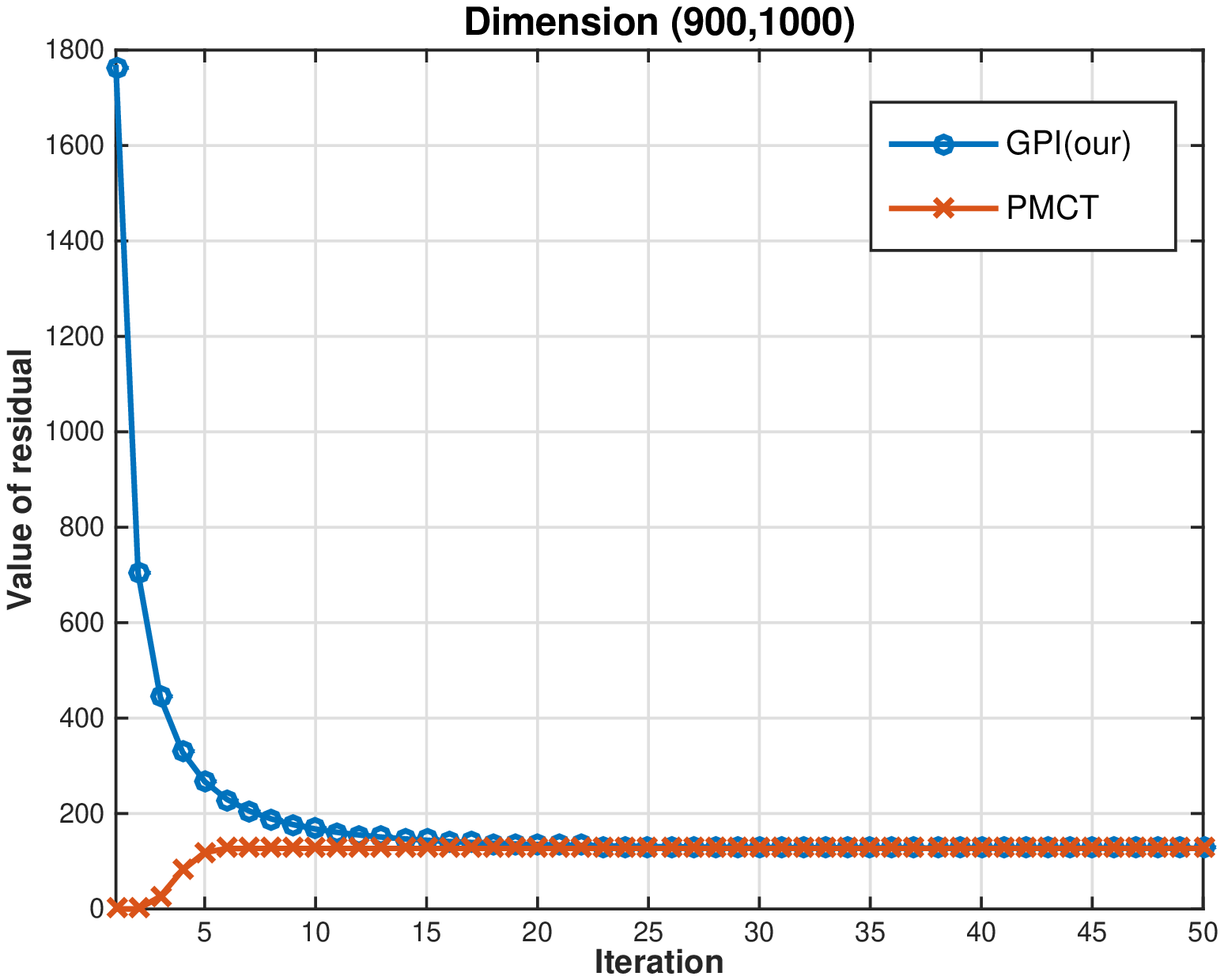}}
    &\subfigure[$(2000,1700)$]{\includegraphics[width=0.5\textwidth,height=0.33\textwidth]{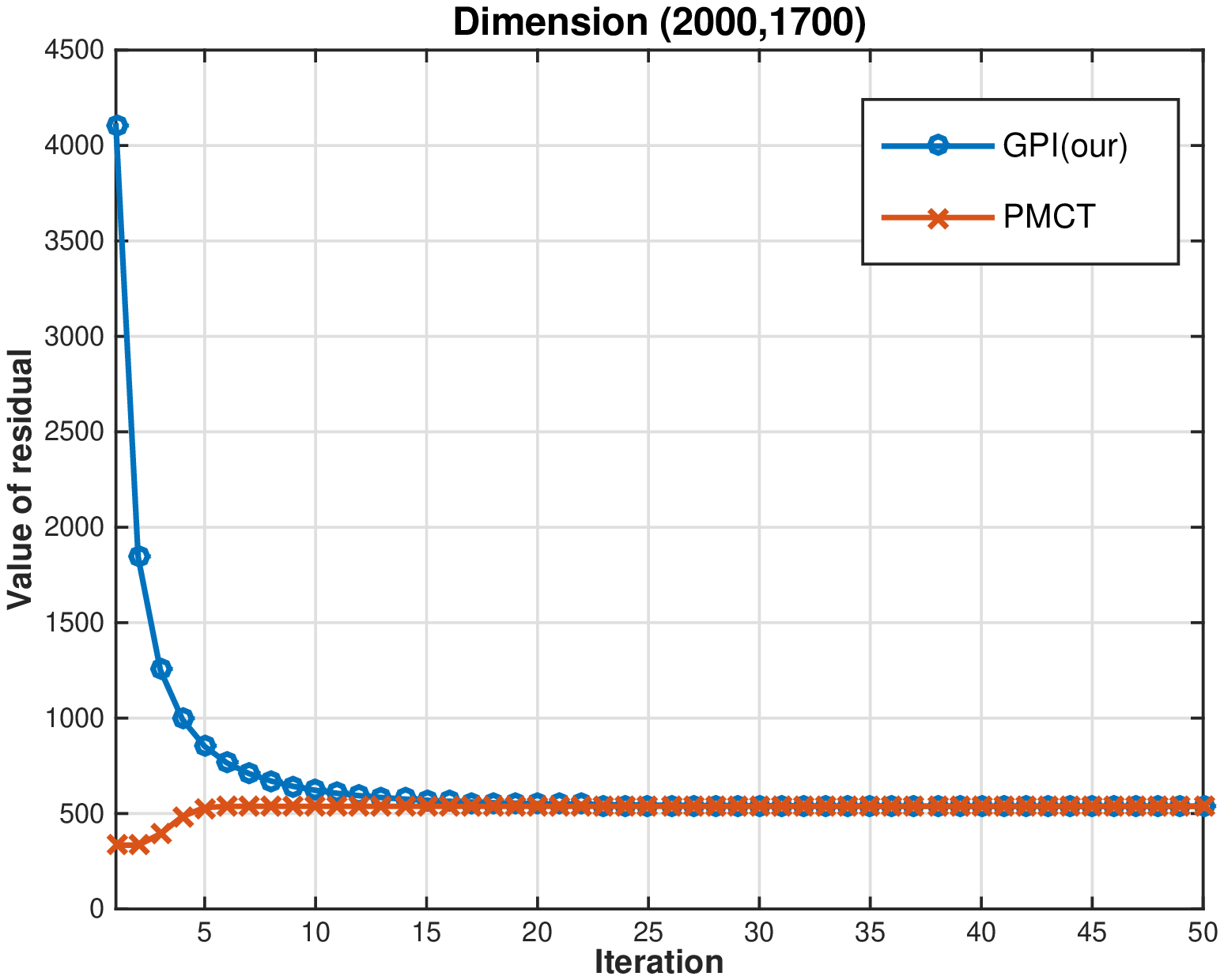}}\\
   \end{tabular}
  \caption{Comparisons of PMCT \cite{Z.H} and GPI are performed over 2 different data matrices.} \label{fig9}
\end{figure*}

\begin{table*}[htbp]
\centering
\caption{Comparisons of CPU time under the square matrix $E$ for Case 2.}\label{tab3}
\begin{tabular}{cccccccc}
\multicolumn{8}{c}{(Iteration stops when $\Vert EQ_{i-1}-G\Vert_F^2-\Vert EQ_i-G\Vert_F^2\leq\tau$ where $ \tau=10^{-3}$.)  }\\
\hline

\hline
\multicolumn{2}{c}{$(n=m=200)$ }&RSR\cite{P}&LSR\cite{B.L}&SP\cite{Z.D}&LR\cite{X.H}&EB\cite{G.G}&GPI(our)  \\ \cline{1-8}
$k=10$&CPU time&64.940s&23.426s&2.386s&0.541s&0.337s&\textbf{0.228s}\\ 
$k=15$&CPU time&136.020s&21.635s&3.221s&1.134s &0.347s&\textbf{0.226s} \\ 
$k=20$&CPU time&229.851s&20.560s&5.054s&1.806s &0.445s&\textbf{0.273s }\\ 
\hline
\hline
\multicolumn{2}{c}{$(n=m=1000)$ }&RSR\cite{P}&LSR\cite{B.L}&SP\cite{Z.D}&LR\cite{X.H}&EB\cite{G.G}&GPI(our)  \\ \cline{1-8}
$k=10$&CPU time&-&842.849s&132.232s&3.869s&11.440s&\textbf{1.290s}\\ 
$k=15$&CPU time&-&851.231s&196.761s&5.180s&12.534s&\textbf{1.434s} \\ 
$k=20$&CPU time&-&860.746s&260.132s&7.700s&12.625s&\textbf{1.575s }\\ 
\hline
\end{tabular}
\end{table*}

Generally speaking, QPSM can not be reformulated into UOPP while UOPP could always be rewritten into QPSM.  Therefore,  the GPI method is more general than other approaches, which can only cope with UOPP. Based on the experimental results involved in the next section, the proposed GPI method takes much less time to converge to the solution of UOPP. 

\section{Experimental Results}

In this section, we analyze and report the numerical results of the generalized power iteration method (GPI) represented by both the algorithm \ref{alg1} and the algorithm \ref{alg2}. We randomly choose the test data matrix with normally distributed singular values. 
\begin{figure}[htb]
    \centering
    \begin{tabular}{m{0.5\textwidth}}
    \subfigure{\includegraphics[width=0.45\textwidth,height=0.25\textwidth]{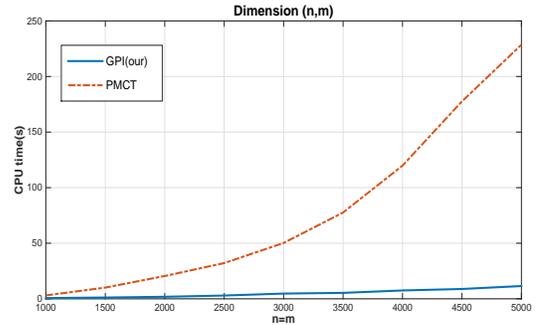}}
\end{tabular}
 \caption{CPU time comparison under Case 3.}\label{fig10}
\end{figure}
Besides, the computer we use is MacBook Air, whose CPU is 1.4 GHz Intel Core i5, RAM is 4 GB 1600 MHz DDR3 and operating system is OS X Yosemite 10.10.5.

  \textit{Case 1: (parameter dependence)} Firstly, we try to investigate the GPI method in the algorithm \ref{alg2} via varying the relaxation parameter $\gamma$. Suppose $l_e$ is the largest eigenvalue of $E^TE$, then we can let $\gamma=\delta l_e$ such that $\gamma I_m-E^TE$ is a positive definite matrix, where $\delta$ is an arbitrary  constant.

1)  From the figure \ref{fig3},  we can further notice that although the convergence rate for the algorithm \ref{alg2} is inversely proportional to the value of $\gamma$,   the relaxation parameter $\gamma$ does not affect the uniform convergence of the GPI method.
\begin{table*}[htbp]
\centering
\caption{Comparison of the CPU time under the general dimension for Case 2.}
\begin{tabular}{cccccccc}
\multicolumn{8}{c}{(Iteration stops when $\Vert EQ_{i-1}-G\Vert_F^2-\Vert EQ_i-G\Vert_F^2\leq\tau$ where $ \tau=10^{-3}$.)  }\\
\hline
Dimension&&\multirow{2}{*}{RSR\cite{P}}&\multirow{2}{*}{LSR\cite{B.L}}&\multirow{2}{*}{SP\cite{Z.D}}&\multirow{2}{*}{LR\cite{X.H}}&\multirow{2}{*}{EB\cite{G.G}}&\multirow{2}{*}{GPI(our)}  \\ 
$(n,m,k)$&&&&&&&\\ \cline{1-8}
\multirow{2}{*}{$(5000,500,15)$}&CPU &\multirow{2}{*}{713.156s}&\multirow{2}{*}{528.034s}&\multirow{2}{*}{450.028s}&\multirow{2}{*}{20.709s}&\multirow{2}{*}{16.554s}&\multirow{2}{*}{\textbf{3.581s}}\\
&time&&&&&&\\ 
\multirow{2}{*}{$(10000,1000,30)$}&CPU &\multirow{2}{*}{-}&\multirow{2}{*}{-}&\multirow{2}{*}{-}&\multirow{2}{*}{56.772s}&\multirow{2}{*}{191.970s}&\multirow{2}{*}{\textbf{9.384s}}\\
&time&&&&&&\\ 
\multirow{2}{*}{$(3000,3000,90)$}&CPU &\multirow{2}{*}{-}&\multirow{2}{*}{-}&\multirow{2}{*}{-}&\multirow{2}{*}{186.125s}&\multirow{2}{*}{395.401s}&\multirow{2}{*}{\textbf{17.320s}}\\
&time&&&&&&\\ 
\multirow{2}{*}{$(30000,1500,30)$}&CPU &\multirow{2}{*}{-}&\multirow{2}{*}{-}&\multirow{2}{*}{-}&\multirow{2}{*}{306.132s}&\multirow{2}{*}{1056.311s}&\multirow{2}{*}{\textbf{19.440s}}\\
&time&&&&&&\\ 
\multirow{2}{*}{$(5000,4000,100)$}&CPU &\multirow{2}{*}{-}&\multirow{2}{*}{-}&\multirow{2}{*}{-}&\multirow{2}{*}{405.937s}&\multirow{2}{*}{1187.512s}&\multirow{2}{*}{\textbf{30.128s}}\\
&time&&&&&&\\ 
\multirow{2}{*}{$(100000,3000,50)$}&CPU &\multirow{2}{*}{-}&\multirow{2}{*}{-}&\multirow{2}{*}{-}&\multirow{2}{*}{-}&\multirow{2}{*}{-}&\multirow{2}{*}{\textbf{215.173s}}\\
&time&&&&&&\\ 
 \cline{1-8}
\end{tabular}\label{tab4}
\end{table*}

 \textit{Case 2:(CPU time comparison for solving UOPP)} Secondly,  we further investigate the proposed GPI method in the algorithm \ref{alg2} by comparing it with five existing approaches mentioned in section 1 as EB \cite{G.G}, RSR \cite{P}, LSR \cite{B.L}, SP \cite{Z.D} and LR \cite{X.H}. 

Based on solving LSQE problem, RSR \cite{P} and LSR \cite{B.L} respectively update the solution row by row and column by column iteratively. EB \cite{G.G} utilizes the expanded balanced OPP as the objective function. SP \cite{Z.D}  employs the projection method combined with correction techniques (PMCT) \cite{Z.H}. LR  \cite{X.H} solves  UOPP by fixing different Lagrangian multipliers.  The proposed GPI method includes two terms as $E^TE$ outside the loop and $\tilde AW$ within the loop, whose orders of complexity are $m^2n$ and $m^2k$, respectively. Besides, these two terms have the highest orders of complexity for the proposed GPI method. Besides, the order of the complexity for each method is shown in the table \ref{tab2.5}.

The comparative results are based on fixing $E$ as the square matrix at first hand (Table \ref{tab3}) and then extend $E$ to a more general case (Table \ref{tab4})  afterwards.   (Mark $-$ in the table \ref{tab3} and the table \ref{tab4} represents that it takes too much time to record in the tables.)

1) From the figure \ref{fig4}, we notice that the existing methods as EB \cite{G.G}, RSR \cite{P}, LSR \cite{B.L}, SP \cite{Z.D} LR \cite{X.H} and the proposed GPI method converge to the same objective value under the same input data. Besides, our proposed GPI method converges faster than other approaches during iteration.

 2) From the table \ref{tab2.5},  the proposed GPI method has the lowest order of complexity due to its succinct computational process to obtain the optimal solution. During the experiments, we observe that the iteration number $t$ for the LR method is usually very large for the convergence. Thus, the time consumption for LR method is much larger than that for the proposed GPI method though orders of complexity for these two approaches seem close.  Besides,  the GPI method becomes more efficient when $n$ (the number of data) is large.

3) From the table \ref{tab3}, the  proposed algorithm \ref{alg2} (GPI) serves as the most efficient method under the square matrix case.

4) From the table \ref{tab4}, we can observe that LSR \cite{B.L}, SP \cite{Z.D}  and RSR \cite{P} are unable to compete with LR \cite{X.H}, EB \cite{G.G} and GPI  due to the complex updating procedures including the expanded OPP and solving LSQE. Especially when the dimension increases, the superiority of our proposed GPI method would be more obvious.

\textit{Case 3:(CPU time comparison for solving LSQE)}  Finally, the projection method combined with correction techniques (PMCT) \cite{Z.H} is compared to the GPI method in the algorithm \ref{alg2}  targeting at solving the least square regression with a quadratic equality constraint (LSQE) in (\ref{ls}). Actually, solving LSQE (\ref{ls}) is no different from solving UOPP (\ref{e1}) under $k=1$. 

1) From the figure \ref{fig9}, we can notice that PMCT \cite{Z.H} and the algorithm \ref{alg2} (GPI) converge to the same objective value though in terms of the different patterns.

2) From the figure \ref{fig10},  the algorithm \ref{alg2} (GPI) takes much less time for convergence than PMCT \cite{Z.H} does. 

\section {Concluding Remarks}
In this paper, we analyze the quadratic problem on the Stiefel manifold (QPSM) by deriving a novel generalized power iteration (GPI) method. Based on the proposed GPI method, two special and significant cases of QPSM known as the orthogonal least square regression and the unbalanced orthogonal procrustes problem are under further investigation.  With the theoretical supports, the GPI method decreases the objective value of the QPSM problem monotonically to a local minimum until convergence. Eventually, the effectiveness and the superiority of the proposed GPI method are verified empirically.  In sum, the proposed  GPI method not only takes less CPU time to converge to the optimal solution with a random initial guess but becomes much more efficient especially for the data matrix of large dimension as well.


\begin{thebibliography}{}
%
%
\bibitem{G}
Green, B.:  The orthogonal approximation of an oblique simple structure
in factor analysis. Psychometrika. {\bf 17} (1952) 429-440. 
\bibitem{H.C}
Hurley, J., Cattell, R.:  The procrustes program: producing direct
rotation to test a hypothesized factor structure. Behavioural Science. {\bf 6} (1962) 258-262. 
\bibitem{G.L}
Golub, G. H., Van Loan, C. F.:  Matrix Computations. The Johns Hopkins
University Press (1989).
\bibitem{T}
Thomas, V.: Algorithms for the weighted orthogonal procrustes problem and other least squares problems: [D]. (2006)  Ume\aa \ University, Sweden. 
\bibitem{R.S.C.R}
Souza, P., Leite, C., Borges, H., Fonseca, R.: Online algorithm based on support vectors for orthogonal regression. Pattern Recognition Letters
{\bf 34} (2013) 1394-1404. 
\bibitem{C.T}
Chu, M., Trendafilov, N.:  The orthogonally constrained regression
revisted.  J. Comput. Graph. Stat. {\bf 10} (2001) 746-771. 
\bibitem{G.G}
Green, B., Goers, J.:  A problem with congruence. The Annual Meeting of the Psychometric Society. Monterey, California (1979).
\bibitem{P}
Park, H.:  A parallel algorithm for the unbalanced orthogonal procrustes problem. Parallel Computing. {\bf 17} (1991)  913-923.
\bibitem{B.L}
Bojanczyk, A., Lutoborski, A.:  The procrustes problem for orthogonal stiefel matrices. SIAM. J. Sci. Comput. {\bf 21} (1999) 1291-1304. 
\bibitem{Z.D}
Zhang, Z., Du, K.:  Successive projection method for solving the unbalanced procrustes problem. Science in China: Series A Mathematics. {\bf 49} (2006) 971-986. 
\bibitem{X.H}
Xia, Y., Han, Y.: Partial lagrangian relaxation for the unbalanced orthogonal procrustes problem. Math. Meth. Oper. Res. {\bf 79} (2014)  225-237. 
\bibitem{Z.H}
Zhang, Z., Huang, Y.:  A projection method for least square problems with quadratic equality constraint. 
SIAM. J. Matr. Anal. Appl. {\bf 25} (2003)188-212. 
\bibitem{MYPR}
Journ\'ee, M., Nesterov, Y., Richt\'arik, P., Sepulchre, R.: Generalized power method for sparse principal component analysis. Journal of Machine Learning Research. {\bf 11} (2008) 517-553.
\bibitem{fiori1}
Fiori, S.:  Formulation and integration of learning differential equations on the Stiefel manifold. IEEE Transactions on Neural Networks. {\bf 16} (2005) 1697-1701.
\bibitem{fiori2}
Kaneko, T., Fiori, S., Tanaka, T.: Empirical arithmetic averaging over the compact stiefel manifold. IEEE Transactions on Signal Processing. {\bf 61} (2013) 883-894.
\bibitem{N.Y.H}
Nie, F., Yuan, J., Huang, H.:  Optimal mean robust principal component analysis. in Proc. ICML. (2014) 2755-2763. 
\end{thebibliography}
\end{document}